%% file: SymFermi.tex
\documentclass[prb,aps,twocolumn, superscriptaddress, nofootinbib]{revtex4-2}

\input{sdpPreamble.tex}
\begin{document}

\hfill MIT-CTP/5947

\title{
Symmetry-Enforced Fermi Surfaces
}

\author{Minho Luke Kim}
\thanks{These authors contributed equally to this work.}
\affiliation{Department of Physics, Massachusetts Institute of Technology}

\author{Salvatore D. Pace}
\thanks{These authors contributed equally to this work.}
\affiliation{Department of Physics, Massachusetts Institute of Technology}

\author{Shu-Heng Shao}
\affiliation{Center for Theoretical Physics --- a Leinweber Institute, Massachusetts Institute of Technology}

\begin{abstract}

We identify a symmetry that enforces every symmetric model to have a Fermi surface. These symmetry-enforced Fermi surfaces are realizations of a powerful form of symmetry-enforced gaplessness. The symmetry we construct exists in quantum lattice fermion models on a $d$-dimensional Bravais lattice, and is generated by the on-site U(1) fermion number symmetry and non-on-site Majorana translation symmetry. The resulting symmetry group is a noncompact Lie group closely related to the Onsager algebra. For a symmetry-enforced Fermi surface $\cF$, we show that this UV symmetry group always includes the subgroup of the ersatz Fermi liquid L$_{\cF}$U(1) symmetry group formed by even functions ${f(\k)\in\mathrm{U}(1)}$ with ${\k\in\cF}$. Furthermore, we comment on the topology of these symmetry-enforced Fermi surfaces, proving they generically exhibit at least two noncontractible components (i.e., open orbits).

\end{abstract}

\maketitle

\makeatletter 
\def\l@subsection#1#2{}
\makeatother 

\tableofcontents

\section{Introduction}

Quantum phases of matter come in two distinct flavors: gapped phases and gapless phases.

A gapped phase is a phase whose energy spectrum in the thermodynamic limit has a finite gap separating the ground state energy from the first excited state energy. The landscape of gapped phases has been heavily explored and is remarkably rich. It includes, for example, discrete symmetry breaking phases, symmetry-protected topological and topologically ordered phases~\cite{W161003911}, and fracton phases~\cite{NH180311196}. 

A gapless phase is a phase whose energy spectrum in the thermodynamic limit does not have a finite gap separating the ground state energy from the first excited state energy. 
The landscape of gapless phases is under much less control compared to that of gapped phases. 
One clear thing, however, is that not all gapless phases are equally gapless. Indeed, gapless phases can be organized by, roughly speaking, the number of gapless excitations---the number of gapless fields in the infrared (IR) effective field theory description of the phase. Some gapless phases have a finite number of gapless excitation types. This includes, for example, conventional ordered phases with Goldstone bosons and critical phases whose IR limit is described by conformal field theories. There are also gapless phases with an infinite number of gapless excitations. A prototypical example of such a gapless phase is a metal, which has an entire Fermi surface worth of gapless excitations.

A fruitful approach to studying gapless phases is to identify ultraviolet (UV) constraints that forbid gapped phases, thereby enforcing gaplessness. This is most commonly realized by imposing anomalous UV symmetries that enforce gaplessness~\cite{WS14011142, WS160406807, SKW160908616, WNM170302426, GHO171004218, CO191004962, CO191213069, ACL221214605, CPS240912220, PCS241218606, PKC250504684,Hsin:2025ria}. We distinguish two notions of this symmetry-enforced gaplessness (SEG): \emph{weak} and \emph{strong} SEG.\footnote{What we call weak SEG is often referred to as just SEG in the literature.} 

Weak SEG arises when an anomalous
symmetry is incompatible with topological and fracton orders, and enforces gaplessness whenever its discrete symmetries are not spontaneously broken. Weak SEG permits gapped, discrete symmetry-broken phases. A trivial example of weak SEG is from anomalous discrete symmetries in ${(1+1)}$D because there is no topological/fracton order in ${(1+1)}$D. One of the first nontrivial examples of weak SEG~\cite{WS14011142} is from a ${(\mathrm{SU}(2)\times\Z_4^T)/\Z_2}$ symmetry in ${(2+1)}$D whose 't Hooft anomaly enforces either gaplessness or a gapped phase with spontaneously broken $\Z_4^T$ time-reversal symmetry.

Strong SEG, on the other hand, occurs when an anomalous symmetry is incompatible with all gapped phases. 
Strong SEG enforces gaplessness regardless of whether its symmetry is spontaneously broken. For example, perturbative anomalies of continuous global symmetries cause strong SEG because they are encoded in the local operator product expansion and, therefore, cannot be matched by any gapped phase~\cite{tHooft:1979rat}. A less trivial example of strong SEG is Witten's SU(2) anomaly in ${(3+1)}$D~\cite{GHO171004218}.  

All known examples of SEG, both of the weak and strong form, enforce gaplessness that is compatible with a finite number of gapless excitations described by a conventional continuum field theory. It is interesting to wonder, however, if there can be an SEG that requires an infinite number of gapless excitations. In this Letter, we answer the following question in the affirmative: is there a symmetry that can enforce a Fermi surface? In particular, we show that quantum lattice models with a single, spinless fermion per unit cell must have a Fermi surface if they (1) conserve the total fermion number and (2) commute with \emph{Majorana} lattice translations. 
This is a powerful form of strong SEG.

\section{Enforcing a Fermi surface}

A Fermi surface is a codimension-1 locus in momentum space where fermionic gapless excitations reside.\footnote{Our definition of Fermi surface relies on momentum space and, therefore, requires translation symmetry. Accordingly, we assume lattice translation symmetry throughout so that crystal momentum is a good quantum number.} We will comment on both ${d=1}$ and ${d>1}$ spatial dimensions throughout the Letter, but primarily focus on ${d>1}$ and only use the term Fermi surface in that case. If a symmetry enforces a Fermi surface, then every microscopic model with that symmetry must have a Fermi surface. Importantly, this symmetry is only a  sufficient condition for a Fermi surface, and it does not mean that every model with a Fermi surface must have this UV symmetry. Given such a symmetry and a symmetric model, the only way to destroy the model's symmetry-enforced Fermi surface would be to explicitly break its UV symmetry.

We investigate the possibility of symmetry-enforced Fermi surfaces within a familiar class of UV models, described by quantum Hamiltonian lattice models. We assume the $d$-dimensional spatial lattice $\La$ has one lattice site per unit cell and denote by $\mb{a}_i$, with ${i=1,2,\cdots, d}$, a choice of its primitive lattice vectors. For instance, in ${d=2}$, $\Lambda$ can be a square lattice but not a honeycomb lattice.
The lattice vectors ${\r = \sum_{i=1}^d n_i \mb{a}_i}$ satisfy the periodic boundary conditions ${\r \sim \r + L_i \mb{a}_i}$ for each $i$, where $L_i$ is a positive integer. We assume the total number of lattice sites ${|\La| = \prod_{i=1}^d L_i}$ is even. The Hilbert space is defined by having a single complex fermion reside at each site ${\r}$. The local Hilbert space on each site is a
two-level system, acted on by the complex fermion operator $c_\r$. These fermion operators satisfy the standard anticommutation relations ${\{c_\r , c_{\r'}^\dag\} = \del_{\r,\r'}}$ and ${\{c_\r , c_{\r'}\} = 0}$. The Hamiltonian $H$ is assumed to be a local Hamiltonian and admits the decomposition ${H = \sum_{\r\in\La} H_\r}$, where $H_\r$ is a bosonic, hermitian operator acting nontrivially only within a finite range about $\r$. Lastly, we consider UV symmetries described by a collection of unitary operators ${\{U_g\mid g\in G\}}$ acting as representations of a group $G$ for the total Hilbert space and commuting with $H$: ${[H, U_g] = 0}$. 

A symmetry that enforces a Fermi surface rules out a wide class of terms in the Hamiltonian $H$. For example, the symmetry must forbid the chemical potential term ${H_{\mu} = -\mu\sum_\r c^\dag_\r c_\r}$. That is, there must be at least one ${g\in G}$ for which ${[H_\mu, U_g] \neq 0}$. If this were not the case, then $H_\mu$ by itself would be an allowed Hamiltonian, but $H_\mu$ is a gapped Hamiltonian and does not have a Fermi surface. A Fermi-surface-enforcing symmetry must also prevent the system from realizing a nontrivial gapped phase. For example, the symmetry must forbid a charge density wave phase by disallowing, e.g., the density-density interaction ${\sum_{\<\r_1,\r_2\>} c^\dag_{\r_1}c_{\r_1}c^\dag_{\r_2}c_{\r_2}}$~\cite{Sachdev_2023, COBV09120646}, where ${\<\r_1,\r_2\>}$ are nearest-neighbor sites. In general, such a symmetry must forbid $H$ from being built only out of number operators $c^\dag_\r c_\r$. Indeed, such a Hamiltonian can be written as ${\sum_\r \prod_{\mb{v}} c^\dag_{\r+\mb{v}}c_{\r+\mb{v}}}$ over some finite collection of vectors $\mb{v}$. This Hamiltonian is exactly solvable by diagonalizing the number operators and has a gapped spectrum without a Fermi surface.

This approach of enforcing a Fermi surface differs from the approach in Ref.~\onlinecite{Else:2020jln}. There, it is argued that any particle number conserving, translation-invariant, compressible system must have a Fermi surface. These constraints are not purely UV constraints, but a mixture of UV and IR constraints. Imposing U(1) particle number and translation symmetries is a UV constraint, but requiring the ground state to be compressible is an IR constraint. Importantly, these UV constraints alone do not enforce a Fermi surface, and, furthermore, are compatible with a trivially gapped symmetric phase.  It is further argued in~Ref.~\onlinecite{Else:2020jln} that their mixed UV-IR constraints imply an infinite-dimensional Lie group symmetry in the IR, e.g., an LU(1) symmetry in the IR. Importantly, this symmetry and its 't Hooft anomaly exist in the IR and are not UV constraints.

\section{Fermi surface from Majorana translations}

We will now present a symmetry in our class of UV models that enforces a Fermi surface. For a $d$-dimensional lattice $\La$, this symmetry is generated by ${(d+1)}$ locality-preserving unitary operators.

The first symmetry we enforce is the U(1) symmetry associated with fermion number conservation. This U(1) symmetry has the conserved charge operator\footnote{We include a shift of ${-1/2}$ in the summand of $Q$ for purely aesthetic reasons. With this shift, the $Q$ operator~\eqref{Q def} is identified with $Q_{\mb{0}}$ defined later in Eq.~\eqref{Qv def}. This shift does not affect the U(1) symmetry transformation~\eqref{U(1)SymTransf}.}
\begin{equation}\label{Q def}
    Q = \sum_{\r\in\La} \left(c_\r^\dag c_\r - \frac12\right).
\end{equation}
Because we assume that the number of lattice sites $|\La|$ is even, $Q$ has integer-quantized eigenvalues. The U(1) symmetry operator $\ee^{\ii\th Q}$ is on site, satisfies ${\ee^{2\pi \ii Q} = 1}$, and generates the symmetry transformation
\begin{equation}\label{U(1)SymTransf}
    \ee^{\ii\th Q} c_\r \ee^{-\ii\th Q} = \ee^{-\ii\th} c_\r.
\end{equation}
This U(1) symmetry is useful to enforce a Fermi surface since it forbids certain Fermi surface destroying terms, such as pairing terms ${\sum_{\<\r_1,\r_2\>} (c_{\r_1}c_{\r_2} + c^\dag_{\r_2}c^\dag_{\r_1})}$.

To motivate the remaining symmetry operators, recall that the total symmetry must forbid the chemical potential term $H_\mu$ to enforce a Fermi surface. A simple unitary operator that does not commute with $H_\mu$ is the $\Z_2$ charge conjugation operator ${\mathsf{C}}$, which satisfies ${\mathsf{C}\, c_\r \mathsf{C}^\dag = c^\dag_\r}$.\footnote{The operator $\mathsf{C}$ is called charge conjugation because it acts on the U(1) symmetry charge operator~\eqref{Q def} as $\mathsf{C} Q \mathsf{C}^\dag = -Q$. This symmetry is sometimes called particle-hole symmetry.} However, $\mathsf{C}$ and $\ee^{\ii\th Q}$ do not enforce a Fermi surface and are, furthermore, anomaly-free.\footnote{A trivially gapped Hamiltonian that commutes with $\mathsf{C}$ and $\ee^{\ii\th Q}$ can be constructed as follows. For each lattice site $\r$, choose another lattice site ${p(\r)\neq \r}$ with ${|\r - p(\r)|\sim\cO(1)}$ and ${p(p(\r)) = \r}$ (recall we assume the number of lattice sites is even, so each $\r$ has a unique $p(\r)$). We denote the pair of lattice sites $\r$ and $p(\r)$ by ${[\r,p(\r)]}$ and consider the Hamiltonian ${\ii\sum_{[\r,p(\r)]} (c^\dag_{\r}c_{p(\r)} + c_{\r}c^\dag_{p(\r)})}$. This Hamiltonian commutes with $\mathsf{C}$ and $\ee^{\ii\th Q}$ and has a unique gapped ground state. Note that this Hamiltonian is not translation-invariant. If lattice translation symmetry is enforced, then $\mathsf{C}$ and $\ee^{\ii\th Q}$ have an LSM anomaly. This LSM anomaly, however, does not enforce a Fermi surface, e.g., it is compatible with a gapped SSB phase.}

While the addition of $\mathsf{C}$ did not enforce a Fermi surface, the way $\mathsf{C}$ failed to commute with $H_\mu$ hints at another operator. Consider the Majorana fermion operators
\begin{equation}\label{a and b Maj ops}
    a_\r = c^\dag_\r + c_\r,
    \quad
    b_\r = \ii (c^\dag_\r - c_\r),
\end{equation}
which satisfy the reality conditions ${a_\r^\dag = a_\r}$ and $b_\r^\dag = b_\r$ as well as the anticommutation relations ${\{a_{\r_1}, b_{\r_2}\} = 0}$, ${\{a_{\r_1}, a_{\r_2}\} = 2\del_{\r_1, \r_2}}$, and ${\{b_{\r_1}, b_{\r_2}\} = 2\del_{\r_1, \r_2}}$. From its action on $c_\r$, $\mathsf{C}$ satisfies ${\mathsf{C}\, a_\r \mathsf{C}^\dag = a_\r}$ and ${\mathsf{C}\, b_\r \mathsf{C}^\dag = -b_\r}$, which makes $\mathsf{C}$ the $b$-Majorana fermion number parity operator. Using the Majorana operators, the chemical potential term is ${H_\mu = -\frac{\mu\ii}{2}\sum_\r a_\r b_\r +\text{constant}}$. Therefore,  $H_\mu$ fails to commute with $\mathsf{C}$ because $\mathsf{C}$ acts differently on the $a$ and $b$ Majorana operators.

Motivated by this, we consider the $b$-Majorana translation operator $T^{(b)}_{\mb{v}}$ by the lattice vector $\mb{v}$. It satisfies
\begin{equation}
    T_{\mb{v}}^{(b)} a_\r (T_{\mb{v}}^{(b)})^{-1} = a_\r, \quad 
    T_{\mb{v}}^{(b)} b_\r (T_{\mb{v}}^{(b)})^{-1} = b_{\r + \mb{v}},
\end{equation}
thus acting differently on the $a$ and $b$ Majorana operators. Its action on the complex fermion operator is\footnote{There is an alternative basis of complex fermions in which $T_{\mb{v}}^{(b)}$ acts simply. They are given by the complex fermions operators ${c^{\,(a)}_\r = \frac12(a_\r - \ii a_{\r + \mb{a}_i})}$ and ${c^{\,(b)}_\r = \frac12(b_\r - \ii b_{\r + \mb{a}_i})}$, where $\mb{a}_i$ is a direction for which $L_i$ is even. Because we assume the number of sites $|\La|$ is even, there is always at least one such $L_i$. The $\ee^{\ii\th Q}$ and $T_{\mb{v}}^{(b)}$ operators act on these complex fermions by
\begin{align}
    \ee^{\ii\th Q} 
    \begin{pmatrix}
    c^{(a)}_\r\\
    c^{(b)}_\r
    \end{pmatrix}
    \ee^{-\ii\th Q} &=
    \begin{pmatrix}
        \cos(\th) & \sin(\th)\\
        -\sin(\th) & \cos(\th)
    \end{pmatrix}
    \begin{pmatrix}
    c^{(a)}_\r\\
    c^{(b)}_\r
    \end{pmatrix},\nonumber\\
    T_{\mb{v}}^{(b)} 
    \begin{pmatrix}
    c^{(a)}_\r\\
    c^{(b)}_\r
    \end{pmatrix}
    (T_{\mb{v}}^{(b)})^\dag &=
    \begin{pmatrix}
    c^{(a)}_\r\\
    c^{(b)}_{\r + \mb v}
    \end{pmatrix}. \nonumber
\end{align}
} 
\begin{equation}\label{maj transl c ops}
    T_{\mb{v}}^{(b)} c_\r (T_{\mb{v}}^{(b)})^{-1} = \frac12(c^\dag_\r + c_\r - c^\dag_{\r+\mb{v}} + c_{\r+\mb{v}}).
\end{equation}
The $b$-Majorana translations are generated by the $d$ locality-preserving, non-on-site unitary operators $T_{\mb{a}_i}^{(b)}$. Enforcing that every $T_{\mb{a}_i}^{(b)}$ operator commutes with $H$ causes every local symmetric Hamiltonian to admit the decomposition ${H = \sum_{\r\in\La} (H_\r(a) + H_\r(b))}$. This decomposition is  due to the locality condition on $H$. Note that $T^{(b)}_{\mb{v}}$ will generically transform a hole state or a particle state to a superposition of hole and particle states.

The $b$-Majorana translation symmetry has a mixed anomaly with the fermion number parity symmetry generated by ${(-1)^F = \ee^{\ii\pi Q}}$.\footnote{See~\cite{RZF150503966, Hsieh:2016emq, SS230702534,Seiberg:2025zqx} for related discussions on  anomalies of Majorana translations.} For periodic boundary conditions on $c_\r$, this anomaly manifests through the projective algebra 
\begin{equation}
    T^{(b)}_{\mb{a}_i}(-1)^F = (-1)^{(L_i -1)\prod_{j\neq i}L_j}(-1)^F\,T^{(b)}_{\mb{a}_i}.
\end{equation}
This anomaly, however, does not enforce a Fermi surface and is compatible with a nontrivial gapped phase (see~\cite{RZF150503966} for an example in ${d=1}$).

What type of local Hamiltonians in our class of UV models commute with $b$-Majorana translations? Consider, for example, the U(1) symmetric, free fermion model whose Hamiltonian is
\begin{equation}
    \sum_{\r_1,\r_2\in\La} t_{\r_1,\r_2} c_{\r_1}^\dag c_{\r_2},
\end{equation}
with ${t = t^\dag}$. For this Hamiltonian to be local, we require the matrix element ${t_{\r_1,\r_2} \neq 0}$ only if ${|\r_1-\r_2|\sim\cO(1)}$. In terms of the Majorana operators~\eqref{a and b Maj ops}, this Hamiltonian is
\begin{align}\label{genU(1)FreeFermionMaj}
    \frac14\sum_{\r_1,\r_2\in\La} \bigg(&
    t_{\r_1,\r_2}(a_{\r_1}a_{\r_2} + b_{\r_1}b_{\r_2})\\
    &+
    \ii (t_{\r_1,\r_2}+t_{\r_2,\r_1})a_{\r_1}b_{\r_2}
    \bigg).\nonumber
\end{align}

For the operators $T_{\mb{a}_i}^{(b)}$ to commute with a local Hamiltonian~\eqref{genU(1)FreeFermionMaj}, the matrix $t$ must satisfy ${t_{\r_1,\r_2} = t_{\r_1-\mb{a}_i,\r_2-\mb{a}_i}}$ and ${t_{\r_1,\r_2} = -t_{\r_2,\r_1}}$. These conditions are satisfied by ${t_{\r_1,\r_2} = \ii g_{\r_2-\r_1}}$ where ${g_{-\r} = -g_{\r}\in\R}$. Therefore, assuming $b$-Majorana translation symmetry, this Hamiltonian can be written as
\begin{equation}\label{Hg Hamiltonian}
\begin{aligned}
    H &= \frac{\ii}4\sum_{\r_1,\r_2\in\La} 
    g_{\r_2- \r_1}(a_{\r_1}a_{\r_2} + b_{\r_1}b_{\r_2}),\\
    & = \ii \sum_{\r_1,\r_2\in\La} g_{\r_2- \r_1} c^\dag_{\r_1} c_{\r_2}.
\end{aligned}
\end{equation}
Note that $H$ does not include a chemical potential term because ${g_{\mb{0}} = 0}$. Furthermore, enforcing $b$-Majorana translations automatically enforces lattice translations for the complex fermions due to locality.

It is a straightforward generalization of the argument from {Ref.~\onlinecite{CPS240912220} to show that the Hamiltonian~\eqref{Hg Hamiltonian} is, in fact, the most general local Hamiltonian commuting with $Q$ and the $b$-Majorana translations $T_{\mb{a}_i}^{(b)}$. Recall that enforcing $T_{\mb{a}_i}^{(b)}$ causes the $a$ and $b$ Majorana operators to decouple, and note that ~\eqref{Hg Hamiltonian} is the most general symmetric, local, quadratic  Hamiltonian. Therefore, it suffices to show that the generic local deformation ${\sum_{\r_i}(t_{\{\r_i\}}^a\prod_{i=1}^{2n} a_{\r_{i}} + t_{\{\r_i\}}^b\prod_{i=1}^{2n} b_{\r_{i}})}$ is not symmetric when ${n>1}$. An infinitesimal U(1) transformation shifts $a_\r$ and $b_\r$ by $\del a_\r = \ii \theta [Q,a_\r]=  \th b_\r$ and ${\del b_\r = \ii \theta [Q,b_\r]=  -\th a_\r}$, respectively. This causes the first (second) term in the generic deformation to get shifted by operators with ${(2n-1)}$ $a$'s ($b$'s) and one $b$ ($a$). These shifts cannot all cancel unless ${n=1}$. Therefore, enforcing $Q$ and $T_{\mb{a}_i}^{(b)}$ to commute with $H$ enforces all local symmetric Hamiltonians to be free fermions with Hamiltonian~\eqref{Hg Hamiltonian}.\footnote{We emphasize that restricting to free fermions was not an input of the UV data. Upon enforcing the symmetry and locality, the Hamiltonian turned out to be non-interacting. There are non-local interacting terms that are symmetric. Each such term can be constructed by taking a U(1) symmetric, interacting Hamiltonian and summing over each $T_{\mb{a}_i}^{(b)}$ orbit.}

The Hamiltonian~\eqref{Hg Hamiltonian} always has a Fermi surface. Indeed, it is a tight-binding model for a Bravais lattice with no pairing or chemical potential terms. This can be seen explicitly using its single-particle dispersion ${\eps_{\k} = -2 \sum_{\mb{v}} g_{\mb{v}} \sin(\k \cdot \mb{v})}$. The Fermi surface is the codimension-1 locus where ${\eps_{\k} = 0}$, which we denote by
\begin{equation}
    \cF = \{\k \mid \eps_\k = 0\}.
\end{equation}
The dispersion satisfies ${\eps_{-\k} = -\eps_{\k}}$. Therefore, the sets of momenta ${\{\k\mid \eps_\k>0\}}$ and ${\{\k\mid \eps_\k<0\}}$ will always get exchanged under the inversion ${\k\mapsto -\k}$. Because these sets are nonempty and $\eps_\k$ is smooth, the zero set of $\eps_\k$, which is $\cF$, will also be nonempty by the intermediate value theorem and generically define a codimension-1 locus. When ${d=1}$, this means there is always a nonzero, finite number of gapless modes. Two of these modes always appear at ${k=0}$ and ${k=\pi}$. When ${d>1}$, this codimension-1 locus defines a hypersurface. The $Q$ and $T^{(b)}_{\mb{v}}$ operators, therefore, generate a symmetry that enforces a Fermi surface.  Furthermore, because ${\eps_\k  =-\eps_{-\k}}$, the volume of the Fermi sea is always half the volume of the Brillouin zone, and this symmetry enforces the ground state's fermion number filling per unit cell to always be ${1/2}$. 

The symmetry represented by $\ee^{\ii\th Q}$ is a U(1) symmetry, and the symmetry represented by $T^{(b)}_{\mb{v}}$ is a $\prod_{i=1}^d\Z_{L_i}$ symmetry. The total symmetry, however, is not a direct product of these two groups. The Majorana translations act on $Q$. Let us define
\begin{align}
    Q_{\mb{v}} &= T^{(b)}_{\mb{v}} Q (T^{(b)}_{\mb{v}})^{-1} = \frac{\ii}2\sum_{\r} a_{\r} b_{\r + \mb{v}},  \label{Qv def}\\
    G_{\mb{v}} &= \frac{\ii}2\sum_{\r} (a_{\r} a_{\r + \mb{v}} - b_{\r} b_{\r + \mb{v}}).
\end{align}
They satisfy the Lie algebra\footnote{The Lie algebra~\eqref{Ons d alg} can be generated by the $2^d$ basis elements ${\{Q_{\sum_{i=1}^d \!n_i \mb a_i} | \, n_i \in  \{0, 1\} \}}$. For example, when ${d=2}$, these elements are ${\{ Q_{\bm{0}}, Q_{\mb a_1}, Q_{\mb a_2}, Q_{\mb a_1 + \mb a_2} \}}$.}
\begin{equation}\label{Ons d alg}
\begin{aligned}
    &[Q_{\r_1}, Q_{\r_2}] = \ii G_{\r_2 - \r_1}, \qquad [G_{\r_1}, G_{\r_2}] = 0, \\
     &[Q_{\r_1}, G_{\r_2}] = 2\ii (Q_{\r_1 - \r_2} - Q_{\r_1 + \r_2}).
\end{aligned}
\end{equation}
While each $Q_{\mb{v}}$ has integer-quantized eigenvalues, the eigenvalues of $G_{\mb{v}}$ do not have any general quantization conditions. This Lie algebra is a generalization of the Onsager algebra~\cite{PhysRev.65.117} introduced in~Ref.~\onlinecite{PKC250504684}. Denoting the corresponding Lie group as $\mathrm{Ons}_{d}$, the total symmetry group generated by $\ee^{\ii\th Q}$ and Majorana translations is ${\mathrm{Ons}_{d} \rtimes \prod_{i=1}^d\Z_{L_i}}$. Note that because $G_{\mb{v}}$ does not have integer-quantized eigenvalues, $\mathrm{Ons}_{d}$ includes $\R$ subgroups and is a noncompact Lie group. Furthermore, in the thermodynamic limit, there is an infinite number of conserved charges and $\mathrm{Ons}_{d}$ is an infinite-dimensional Lie group.

As symmetries of lattice fermion models, the group $\mathrm{Ons}_{1}$ has appeared in~\cite{VOF181209091,CPS240912220, X250110837, GT250307708, PCS241218606} and $\mathrm{Ons}_{2}$ in~\cite{PKC250504684}. For example, the model~\eqref{Hg Hamiltonian} in ${(1+1)}$D with only nearest neighbor hopping flows to a massless free Dirac fermion in the IR, and the $\mathrm{Ons}_{1}$ symmetry becomes the ${\mathrm{U}(1)_\text{L}\times \mathrm{U}(1)_\text{R}}$ symmetry of the Dirac fermion. In fact, the chiral anomaly of a Dirac fermion is matched by the lattice $\mathrm{Ons}_{1}$ symmetry~\cite{CPS240912220}.

As we mentioned earlier, a symmetry that enforces a Fermi surface is a sufficient but not necessary condition for a Fermi surface to exist. For instance, adding a chemical potential term ${-\mu\sum_{\r\in\La} c_\r^\dag c_\r}$ to~\eqref{Hg Hamiltonian} breaks the Majorana translation symmetry but does not destroy the Fermi surface if $\mu$ is small. Another example is the Hamiltonian
\begin{equation}\label{counter ex}
    \sum_{\r\in\La} \sum_{i=1}^d  [\ii (c_\r^\dagger c_{\r+\mb a_i} - c_{\r+\mb a_i}^\dagger c_{\r}) \\-t (c_\r^\dagger c_{\r+\mb a_i} + c_{\r+\mb a_i}^\dagger c_{\r})].
\end{equation}
Denoting by ${\phi = \arctan(t)}$, this Hamiltonian's single-particle dispersion is ${-2\sqrt{1+t^2} \sum_{i=1}^d \sin (\k \cdot \mb a_i + \phi)}$. It has a Fermi surface for all $t$ and is half filled. Even then, this Fermi surface is not symmetry-enforced since the Majorana translations~\eqref{maj transl c ops} commute with~\eqref{counter ex} only when ${t=0}$. 

\subsection{Fermi surface topology}

The Fermi surface of the general symmetric model~\eqref{Hg Hamiltonian} is not arbitrary. It is constrained by the single-particle dispersion satisfying ${\eps_{-\k} = -\eps_\k}$. One consequence of this is that every Fermi surface $\cF$ must pass through ${\k = 0}$ and other inversion-invariant points, and be invariant under the inversion ${\k\mapsto -\k}$. 

The symmetry properties of $\eps_{\k}$ also affect the topology of allowed Fermi surfaces. Consider a generic Fermi surface $\cF$, for which ${\grad\eps_\k\neq 0}$ for all ${\k\in\cF}$, and its connected-component decomposition
\begin{equation}
    \cF = \bigsqcup_{i\in \pi_0(\cF)} \cF^{(i)}.
\end{equation}
Each $\cF^{(i)}$ is a path-connected component of the Fermi surface. In Appendix~\ref{TopologyAppendix}, we prove the following statements about generic Fermi surfaces $\cF$ whose single-particle dispersion satisfies ${\eps_{-\k} = -\eps_\k}$:
\begin{enumerate}
    \item The number of contractible components  of $\cF$ can be zero or nonzero, but is always even. Furthermore, contractible components cannot pass through a point $\k$ invariant under ${\k\mapsto -\k}$. These are points $\k$ such that $2\k$ is a reciprocal lattice vector.
    \item A generic symmetry-enforced Fermi surface always has at least two non-contractible components. In other words, there are open orbits in the Brillouin zone $d$-torus. 
    \item Every point $\k$ invariant under ${\k\mapsto -\k}$ lies on a non-contractible component  of $\cF$.
\end{enumerate}
In Fig.~\ref{fig:fs}, we show examples of symmetry-enforced Fermi surfaces for ${d=2}$ that highlight these topological properties.

\begin{figure}[t!]
\centering
    \includegraphics[width=.48\textwidth]{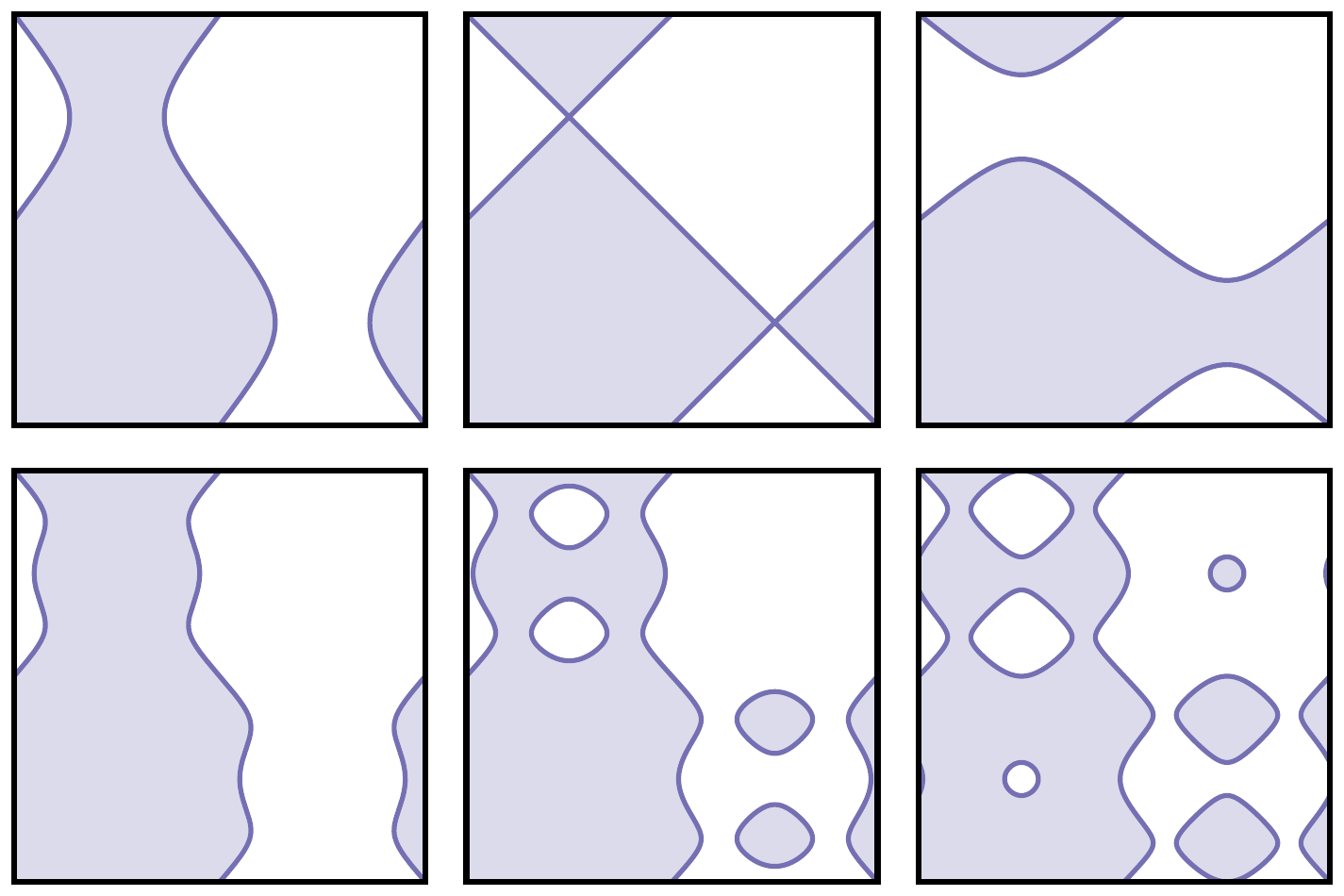}
    \caption{Examples of the Fermi seas and Fermi surfaces enforced by our microscopic symmetry. Specifically, here we consider a special class of Hamiltonians of the form~\eqref{Hg Hamiltonian} on a  ${d=2}$ square lattice with ${\eps_\k = \sin(k_x) + \al \sin(k_y) + \bt (\sin(3k_x)\!+\sin(3k_y))}$. Each panel shows the Brillouin zone with horizontal axis ${-\pi\leq k_x< \pi}$ and vertical axis ${-\pi\leq k_y< \pi}$. The top row shows, from left to right, ${\left(\al,\bt\right) = \left(\frac34,0\right), \left(1,0\right), \left(\frac54,0\right)}$. The bottom row shows, from left to right, ${\left(\al,\bt\right) = \left(\frac34,\frac14\right), \left(\frac34,\frac12\right), \left(\frac34,1\right)}$. Note that the Fermi surface at ${\left(\al,\bt\right) = \left(1,0\right)}$ is nongeneric.
}
    \label{fig:fs}
\end{figure}

\section{Emanant symmetries}

Since the general symmetric Hamiltonian~\eqref{Hg Hamiltonian} always has a Fermi surface, its IR limit does not admit a conventional quantum field theory description (see Refs.~\onlinecite{P9210046, S9307009, Haldane1994, DDM220305004} for a survey of effective field theory approaches). Independent of the formulation of the IR theory, however, we can still investigate what the UV ${\mathrm{Ons}_{d} \rtimes \prod_{i=1}^d\Z_{L_i}}$ symmetry becomes in the IR. In other words, using the terminology introduced in Ref.~\onlinecite{CS221112543}, we can still find the IR symmetry that emanates from the UV ${\mathrm{Ons}_{d} \rtimes \prod_{i=1}^d\Z_{L_i}}$ symmetry. 

The operators acting within the IR are most naturally described in momentum space, where the complex fermion operators become ${c_{\k} = \frac1{\sqrt{|\La|}} \sum_{\r\in\La} \ee^{-\ii \k\cdot \r} c_{\r}}$.\footnote{We follow the convention where $c_\k^\dag$ is the Hermitian conjugate of $c_{\k}$, and not the Fourier transformation of $c_\r^\dag$. In particular, we have ${c^\dag_{\k} = \frac1{\sqrt{|\La|}} \sum_{\r\in\La} \ee^{\ii \k\cdot \r} c^\dag_{\r}}$. Therefore, $c_\k$ ($c^\dag_\k$) creates a hole (particle) at $\k$ with crystal momentum $\k$ ($-\k$).} Because of the spatial periodic boundary conditions, the momentum vector ${\k = \sum_{i=1}^d \frac{m_i}{L_i}\mb{b}_i}$, where ${m_i\in\Z}$ and $\mb{b}_j$ are primitive lattice vectors of the reciprocal lattice $\La^\vee$ which satisfy ${\mb{a}_i\cdot \mb{b}_j = 2\pi\del_{ij}}$. 

The IR limit is achieved by restricting to the low-energy operators and states, and then taking the thermodynamic limit. These low-energy operators are formed by all ${(c_{\k},c_{\k}^\dag)}$ with ${\k\in \cF}$. The emanant symmetry group is the group that faithfully describes the ${\mathrm{Ons}_{d} \rtimes \prod_{i=1}^d\Z_{L_i}}$ symmetry transformations when restricted to operators ${(c_{\k},c_{\k}^\dag)}$ with ${\k\in \cF}$. The emanant symmetry group is, therefore, always isomorphic to a quotient group of ${\mathrm{Ons}_{d} \rtimes \prod_{i=1}^d\Z_{L_i}}$.\footnote{For a UV symmetry group $G_{\mathrm{UV}}$, IR symmetry group $G_{\mathrm{IR}}$, and group homomorphism ${\rho\colon G_{\mathrm{UV}} \to G_{\mathrm{IR}}}$ relating the two, the emanant symmetry group is the image of $\rho$: ${\mathrm{im}(\rho)\equiv\rho(G_{\mathrm{UV}})\subset G_{\mathrm{IR}}}$. By the first isomorphism theorem, ${\mathrm{im}(\rho) \cong G_{\mathrm{UV}}/\ker(\rho)}$. Therefore, the emanant symmetry group is always isomorphic to the quotient group ${G_{\mathrm{UV}}/\ker(\rho)}$. The kernel $\ker(\rho)$ of $\rho$ is the subgroup of $G_{\mathrm{UV}}$ that acts trivially in the IR.}

The Majorana translation symmetry generators emanate to generators of an internal discrete symmetry. The IR limit of the Ons$_d$ symmetry is more exotic. In particular, for ${d>1}$, the Ons$_d$ symmetry emanates to a noncompact, nonabelian, infinite-dimensional Lie group symmetry in the IR.

The IR limit of Ons$_d$ is quite complicated, but it has an interesting subgroup that forms a compact, nonabelian, infinite-dimensional Lie group. Let $\bz^+$ denote a subset of the first Brillouin zone containing only one momentum vector from each pair ${(\k,-\k)}$. In Appendix~\ref{momentum space ons d appendix}, we show that Ons$_d$ has a subgroup isomorphic to
\begin{equation}\label{Ok symmetry UV}
    \prod_{\k\in\bz^+} \!O_\k,
    \qquad
    O_\k = \begin{cases}
        \mathrm{U}(1)\quad &\text{if }2\k\in\La^\vee,\\
        \mathrm{SU}(2)\quad &\text{otherwise.}
    \end{cases}
\end{equation}
The $O_\k$ symmetry with $\k\in\bz^+$ acts nontrivially on only the fermion operators $c_\k$, $c^\dagger_\k$, $c_{-\k}$, and $c^\dagger_{-\k}$. The corresponding conserved charges for $O_\k$ are\footnote{The operator $Q^{(c)}_\k$ can be written as a sum over the charges ${\t{Q}_{\mb{v}} = \frac12(Q_{\mb{v}} + Q_{-\mb{v}})}$. These operators $\t{Q}_{\mb{v}}$ are mutually commuting, and we show in Appendix~\ref{max abl alg app} that they form the maximal abelian sub-algebra of the Ons$_d$ Lie algebra that contains $Q$.}
\begin{align}
    Q^{(c)}_\k &= 
    \frac{2}{|\La|}\sum_{\mb{v}\in\La} \cos(\mb{k}\cdot\mb{v}) \, Q_{\mb{v}},\\
    Q^{(s)}_\k &= \frac{2}{|\La|}\sum_{\mb{v}\in\La} \sin(\mb{k}\cdot\mb{v}) \,Q_{\mb{v}}.
\end{align}
They satisfy ${[Q^{(c)}_\k, c^\dag_{ \k'}] = (\del_{\k,\k'}+\del_{\k,-\k'})c_{\k'}^\dag}$ and ${[Q^{(s)}_\k, c^\dag_{\k'}] = \ii(\del_{\k,\k'}-\del_{\k,-\k'})c_{-\k'}}$. The UV symmetry~\eqref{Ok symmetry UV} becomes a ${\prod_{\k\in\cF^+} \!O_\k}$ symmetry in the IR, where ${\cF^+ \equiv \cF\cap \bz^+}$. This follows from the fact that  if ${\k\in\cF}$, then ${-\k\in\cF}$ too because ${\eps_{-\k} = -\eps_\k}$. 

Let us compare this emanant IR symmetry to the L$_\cF$U(1) symmetry of a Fermi surface $\cF$~\cite{Else:2020jln, ES201010523, DDM220305004, Shi:2022maf, E230110775, LWY230212731, Huang:2024uap} (see~Ref.~\onlinecite{Else:2025hxo} for a recent review). It is conjectured that for every Fermi surface $\cF$, there is an anomalous L$_\cF$U(1) symmetry in the IR, which generically arises as an emergent symmetry. Its symmetry operators are ${\exp[ \ii\it\int_{\k\in\cF} f_\k \, n_\k]}$, where ${f_\k \sim f_\k + 2\pi}$ and $n_\k$ is the fermion density operator, and its symmetry group is the infinite-dimensional Lie group ${\mathrm{L}_\cF\mathrm{U}(1) \equiv \mathrm{Map}[\cF,\mathrm{U}(1)]}$.\footnote{It is an open problem as to which maps from the Fermi surface $\cF$ to U(1) should be included in $\mathrm{L}_\cF\mathrm{U}(1)$. The maps should be at least continuous maps when ${d>1}$ to define the symmetry's 't Hooft anomaly~\cite{Else:2025hxo}.}

The ${\mathrm{Ons}_{d} \rtimes \prod_{i=1}^d\Z_{L_i}}$ symmetry group does not become $\mathrm{L}_\cF\mathrm{U}(1)$ in the IR. Indeed, its emanant symmetries are described by a nonabelian group, while $\mathrm{L}_\cF\mathrm{U}(1)$ is abelian. The  ${\mathrm{Ons}_{d} \rtimes \prod_{i=1}^d\Z_{L_i}}$ symmetry does, however, include a subgroup of ${\mathrm{L}_\cF\mathrm{U}(1)}$. Its symmetry transformation is
\begin{equation}\label{even LU(1) transf}
    \bigg(\!\!\!\!\!\prod_{\,\,\,\,\,\,\k\in\cF^+}\!\!\!\!\!\ee^{\ii f^e_\k Q_\k^{(c)}}\!\bigg)
    \,c_\k\,
    \bigg(\!\!\!\!\!\prod_{\,\,\,\,\,\,\k\in\cF^+}\!\!\!\!\!\ee^{-\ii f^e_\k Q_\k^{(c)}}\!\bigg) \!= 
    \begin{cases}
        \ee^{-\ii f^e_\k} c_\k \qquad &\k\in\cF,\\
        c_\k\qquad&\text{else},
    \end{cases}
\end{equation}
where ${f^e_\k \sim f^e_{\k}+2\pi}$ and ${f^e_\k = f^e_{-\k}}$. This transformation leaves the general symmetric Hamiltonian~\eqref{Hg Hamiltonian} unchanged and is described by the group ${\prod_{\k\in\cF^+}\mathrm{U}(1)}\subset \mathrm{Ons}_{d}$, which we denote as $\mathrm{L}^e_\cF\mathrm{U}(1)$. This is the subgroup of $\mathrm{L}_\cF\mathrm{U}(1)$ made of all even functions ${f^e\colon \cF\to \mathrm{U}(1)}$. It exists only when $\cF$ is inversion symmetric, which is always the case for our symmetry-enforced Fermi surfaces. The transformation~\eqref{even LU(1) transf} acts nontrivially only on IR operators and, therefore, is unchanged upon restricting to the IR. Consequently, a part of the UV ${\mathrm{Ons}_{d} \rtimes \prod_{i=1}^d\Z_{L_i}}$ symmetry becomes $\mathrm{L}^e_\cF\mathrm{U}(1)$ in the IR.

Viewing~\eqref{even LU(1) transf} as a symmetry transformation, the corresponding UV $\mathrm{L}^e_\cF\mathrm{U}(1)$ symmetry is anomaly-free---it is compatible with a trivial, symmetric gapped phase. For example, the symmetry transformation~\eqref{even LU(1) transf} leaves the chemical potential term unchanged. Therefore, the IR anomaly of ${\mathrm{L}_{\cF}\mathrm{U}(1)}$ does not arise from this UV $\mathrm{L}^e_\cF\mathrm{U}(1)$ symmetry.\footnote{We note that IR $\mathrm{L}^e_\cF\mathrm{U}(1)$ symmetry is anomaly-free in ${d=2}$. Indeed, the anomaly of $\mathrm{L}_\cF\mathrm{U}(1)$ manifests by the commutation relation ${[n_\k, n_{\k'}] = -\frac{\ii}{2\pi}\del^{'}(\k - \k')}$ that arises in the presence of a total $2\pi$ U(1) flux~\cite{Else:2020jln}. For $\mathrm{L}^e_\cF\mathrm{U}(1)$, the density $n_\k$ is replaced by $n^e_\k = n_\k+n_{-\k}$. This satisfies ${[n^e_\k, n^e_{\k'}] = 0}$ in the presence of a total $2\pi$ U(1) flux, signaling that $\mathrm{L}^e_\cF\mathrm{U}(1)$ is anomaly-free in the IR. We thank Dominic Else, Marvin Qi, and Zhengyan Darius Shi for related discussion on this point.} However, this $\mathrm{L}^e_\cF\mathrm{U}(1)$ transformation is not locality preserving in real space because its symmetry transformation is localized about $\cF$ in momentum space. It, therefore, is best not to consider~\eqref{even LU(1) transf} as a standalone UV symmetry transformation. For instance, many typical features of anomaly-matching are not expected to hold for such badly locality-violating, conserved operators.\footnote{For example, the operator ${\sum_{j=1}^L \left(\frac{2}{L} \sum_{\pi/2<k<3\pi/2}\cos(j k)\,\right) Q_j}$ in ${d=1}$ generates a U(1) transformation ${c_k\mapsto \ee^{-\ii\th} c_k}$ for ${\pi/2<k<3\pi/2}$ and ${c_k\mapsto  c_k}$ otherwise. This is a highly non-local symmetry of the $({1+1})$D staggered fermion model. In the model's IR limit, it becomes the U$(1)_\text{R}$ chiral symmetry that only acts on the right movers of a free, massless Dirac fermion field theory. While U$(1)_\text{R}$ has the conventional chiral anomaly, the lattice UV symmetry is anomaly-free because it commutes with the chemical potential term.}

\section{Outlook}

In this Letter, we presented a UV symmetry that enforces every local Hamiltonian with that symmetry to have a Fermi surface. The UV symmetry in ${(d+1)}$D was generated by ${d+1}$ locality preserving unitary operators: an on-site fermion number symmetry and a non-on-site Majorana translation symmetry. The resulting symmetry group in the thermodynamic limit is a nonabelian, infinite-dimensional Lie group symmetry. We showed that a generic Fermi surface enforced by this symmetry is always topologically nontrivial.

The UV symmetry included an Onsager-type symmetry. Such symmetries described by Onsager algebras have seen broad interest recently in both fermionic~\cite{VOF181209091,CPS240912220, X250110837, CPS250110862, N250220815, GT250307708, Y250410263, PKC250504684, OY250904906, AKT251106198} and bosonic~\cite{VOF181209091, 2019arXiv190802767O,M210314569, Jones:2024zhx,PCS241218606, S250301831} quantum lattice models. A common theme is that many of these UV Onsager symmetries exhibit new lattice anomalies, often leading to strong SEG. Indeed, they have been used to realize various quantum field theory anomalies on the lattice through anomaly matching, such as the chiral anomaly~\cite{CPS240912220} and the parity anomaly~\cite{PKC250504684}. Despite Onsager-type symmetries often matching quantum field theory anomalies, we did not identify an IR anomaly matched by the UV symmetry that enforces a Fermi surface. It would be interesting to construct a UV symmetry that matches the IR $\mathrm{L}_\cF\mathrm{U}(1)$ anomaly. 

Four other interesting follow-up questions arising from this work are: (1) is there a symmetry that enforces codimension-$p$ Fermi surfaces in ${d>p}$ spatial dimensions? In this Letter, we considered only codimension-1 Fermi surfaces, but higher codimension Fermi surfaces can also arise. For example, codimension-2 Fermi surfaces in ${d=3}$ are reasonably common and called nodal lines. (2) Are there physical consequences of the nontrivial topology of our symmetry-enforced Fermi surfaces similar to those in, for example, Refs.~\onlinecite{TCK220406559,TK221008048,TK231003737}? (3) How are the results presented here affected by turning on a background gauge field of the U(1) fermion number symmetry? This is particularly interesting for a background field corresponding to a uniform magnetic field. We present some preliminary results for the square lattice in Appendix~\ref{magnetic app}. (4) Do symmetric Fermi surfaces have an emergent symmetry larger than $\mathrm{L}_\cF\mathrm{U}(1)$? The Fermi surfaces in this Letter were inversion-symmetric, and the IR symmetry was larger than $\mathrm{L}_\cF\mathrm{U}(1)$. It would be interesting to investigate whether it is a general feature that a symmetric $\cF$ leads to an enhanced $\mathrm{L}_\cF\mathrm{U}(1)$ symmetry.

\section*{Acknowledgments}

We would like to thank {\"O}mer Mert Aksoy, Arkya Chatterjee, Luca Delacr{\'e}taz, Patrick Ledwith, Umang Mehta, Nathan Seiberg, T. Senthil, Xiao-Gang Wen, Wucheng Zhang, and Zhipu Wilson Zhao for interesting discussions. 
In particular, we would like to thank Arkya Chatterjee for collaboration during the early stage of this project. 
We further thank Arkya Chatterjee, Dominic Else, Marvin Qi, and Zhengyan Darius Shi for valuable feedback on the manuscript.
M.L.K. and S.D.P. are supported by the Simons Collaboration on Ultra-Quantum Matter, which is a grant from the Simons Foundation (651446, XGW). M.L.K. is also partially supported by the NSF grant DMR-2022428 (Wen). S.H.S. is supported by the Simons Collaboration on Ultra-Quantum Matter, which is a grant from the Simons Foundation (651444, SHS), and by
NSF grant PHY-2449936.

\appendix

\section{The topology of symmetry-enforced Fermi surfaces}\label{TopologyAppendix}

In this appendix, we prove the statements  in the main text regarding the topology of Fermi surfaces enforced by our microscopic symmetry. We work with a general $d$-dimensional Bravais lattice in the thermodynamic limit. We assume ${d\geq 2}$. The Brillouin zone is a $d$-torus $T^d$. The Fermi surface $\cF$ is a codimension-1 locus in $T^d$ defined by the zero set of the single-particle dispersion $\eps_\k$:
\begin{equation}\label{FS def App}
    \cF = \{ \k \in T^d \mid \eps_\k = 0\}.
\end{equation}
We will often invoke the connected-component decomposition
\begin{equation}\label{con comp decomp}
    \cF = \bigsqcup_{i\in \pi_0(\cF)} \cF^{(i)},
\end{equation}
where each $\cF^{(i)}$ is a path-connected component of the Fermi surface.

Since the Fermi surface $\cF$ is the boundary of the Fermi sea, it is always homologically trivial: ${[\cF] = 0 \in H_{d-1}(T^d)}$. Our microscopic symmetry, generated by the on-site U(1) symmetry and the Majorana translations, forces $\epsilon_\k$ to be a smooth function of $\k$ that satisfies
\begin{equation}\label{inversion}
\epsilon_{-\k } = - \epsilon_\k \,.
\end{equation}
This causes the Fermi surface to be invariant under the central inversion ${I\colon \k\mapsto-\k}$ and contain every point invariant under $I$, e.g., ${\k = \bm{0}}$. In what follows, we will show how these properties constrain the topology of $\cF^{(i)}$ for symmetry-enforced Fermi surfaces $\cF$.

We first note that the symmetry-enforced Fermi surface $\cF$ can have a component $\cF^{(i)}$ that is contractible. Indeed, consider the dispersion
\begin{align}
    \epsilon_{\k}(g) = \sum_{i=1}^d \left(\sin (\k \cdot \mb a_i) + g \sin (3\k \cdot \mb a_i)\right),
\end{align}
which satisfies ${\epsilon_{-\k}(g) = -\epsilon_{\k}(g)}$. Denoting by ${\phi_i = \k\cdot\mb a_i}$, we note that
\begin{equation}\label{derives}
\begin{aligned}
    &\pp_{\phi_i} \epsilon_{\k}(g) = 
       \cos (\k \cdot \mb a_i)  + 3g \cos (3\k \cdot \mb a_i),\\
    &\pp_{\phi_i}\pp_{\phi_j}  \eps_{\k}(g) = -\del_{ij}\left(\sin(\k \cdot \mb a_i)  + 9g \sin(3\k \cdot \mb a_i) \right).
\end{aligned}
\end{equation}
At momentum ${\p = \frac14\sum_{i=1}^d \mb b_i}$,  ${\pp_{\phi_i} \epsilon_{\p}(g) = 0}$ and the Hessian matrix whose elements are $\pp_{\phi_i}\pp_{\phi_j}  \eps_{\k}(g)$ is positive definite for ${g > \frac{1}{9}}$. Hence, $\p$ is a local minimum of $\eps_\k$ if ${g>\frac{1}{9}}$, at which $\epsilon_\p (g) = d (1-g)$. At ${g=1}$, the only point in a neighborhood of $\p$ for which ${\epsilon_\k = 0}$ is $\p$. Because ${\epsilon_\p < 0}$ for finite ${g>1}$ and $\eps_{\k}(g)$ is continuous in $g$, there must be a contractible hypersurface $\cF^{(i)}$ where ${\eps_{\k}(g) = 0}$ in the neighborhood of $\p$ when ${g>1}$.

While a component $\cF^{(i)}$  of $\cF$ can be contractible, the constraint~\eqref{inversion} causes a generic symmetry-enforced Fermi surface always to have at least two homologically nontrivial components. Before proving this, we will first clarify what we mean by \textit{generic} Fermi surface. 

A generic Fermi surface $\cF$ is a Fermi surface for which $\nabla\eps_\k \neq \bm{0}$ for all ${\k \in \cF}$.\footnote{Non-generic Fermi surfaces are measure zero in the space of allowed Fermi surfaces, forming a codimension ${\geq 1}$ subset of parameter space.} Note that if $\cF$ is generic, then each $\cF^{(i)}$ in~\eqref{con comp decomp} is also generic. An important property of a generic Fermi surface is that it cannot have cusps or self-intersections. Indeed, to show this, consider a neighborhood of an arbitrary point ${\mb{q}\in\cF}$. By the implicit function theorem, a generic Fermi surface has a unique differentiable function ${\varphi_i(q_1,\cdots q_{i-1}, q_{i+1}, \cdots q_{d})}$ for each $i$ such that ${\eps_{(q_1,\cdots q_{i-1}, \varphi_i, q_{i+1}, \cdots q_{d})} = 0}$ in a neighborhood of $(q_1,\cdots q_{i-1}, q_{i+1}, \cdots q_{d})$. Note that (1) if a generic Fermi surface had cusps, then $\varphi_i$ is necessarily not differentiable and (2) if a generic Fermi surface had self-intersections, then $\varphi_i$ is necessarily not unique. However, $\varphi_i$ is a unique, differentiable function. Therefore, a generic Fermi surface cannot have cusps or self-intersections.

Let us now discuss the topology of a generic symmetry-enforced Fermi surface $\cF$. The primary result, from which other properties follow, is that every central inversion-symmetric point on a generic symmetry-enforced $\cF$ is on a non-contractible component $\cF^{(i)}$ of $\cF$. To prove this, consider a point ${\p\in\cF^{(i)}}$ that is invariant under ${I\colon \k\mapsto -\k}$. In an effort to prove by contradiction, assume the component $\cF^{(i)}$ is contractible. Since $\cF$ is generic and invariant under ${I}$, the component $\cF^{(i)}$ must also be invariant under $I$ since it contains $\p$. Every contractible, central inversion-symmetric hypersurface that passes through its inversion center has a self-intersection at the inversion center.\footnote{\label{cont footnote}To prove this, consider a contractible hypersurface $S$ with interior $D$ and exterior $E$. Under a central inversion about a point on $S$, $D$ gets mapped to $D$ and $E$ gets mapped to $E$ since $S$ is central inversion symmetric. We denote the inward-pointing normal vector of $S$ at the inversion-center by $\bm{n}$. Suppose $S$ has no self-intersection at the inversion-center. Then ${\eps\,\bm{n}\in D}$ and ${-\eps\,\bm{n} \in E}$ for ${0<\eps\ll 1}$. However, under the central inversion, ${\bm{n}\mapsto -\bm{n}}$. Since $D$ gets mapped to $D$ under central inversion, ${-\eps\,\bm{n} \in D}$. This, however, contradicts the necessary condition that ${-\eps\,\bm{n} \in E}$ if $S$ has no self-intersections at the inversion-center. Therefore, such $S$ must have a self-intersection at the inversion-center.} Therefore, $\cF^{(i)}$ necessarily has a self-intersection at ${\k=\p}$ because it is contractible. However, because $\cF$ is generic, $\cF^{(i)}$ cannot have a self-intersection. Therefore, the initial assumption is wrong, and the component $\cF^{(i)}$ must be non-contractible.

A consequence of this is that every generic symmetry-enforced Fermi surface $\cF$ must have at least two homologically nontrivial components $\cF^{(i)}$. Indeed, every generic Fermi surface $\cF$ with dispersion satisfying ${\eps_{-\k} = - \eps_{\k}}$ has a component $\cF^{(0)}$ passing through ${\k = \bm{0}}$. Because $\bm{0}$ is invariant under ${\k\mapsto -\k}$, the component $\cF^{(0)}$ is always non-contractible. Furthermore, because the total Fermi surface is homologically trivial, there must be at least one component in addition to $\cF^{(0)}$ that is non-contractible to ensure the total Fermi surface is homologically trivial. Thus, there are always at least two non-contractible components of a generic, symmetry-enforced $\cF$.

Another consequence is that the contractible components of a generic symmetry-enforced Fermi surface cannot pass through ${\k\mapsto -\k}$ invariant points. Indeed, as explained in footnote~\ref{cont footnote}, such contractible components would necessarily have self-intersections. However, a generic Fermi surface has no self-intersections, so its contractible components cannot pass through ${\k\mapsto -\k}$ invariant points.

As a result, a generic symmetry-enforced Fermi surface always has an even number of contractible components. Indeed, since the Fermi surface $\cF$ is invariant under ${I\colon \k\mapsto -\k}$, the image $I(\cF^{(i)})$ of each component $\cF^{(i)}$ under $I$ is also a component of $\cF$. Because $\cF$ is generic, a contractible $\cF^{(i)}$ does not pass through an $I$ invariant point. Therefore, a contractible $\cF^{(i)}$ cannot be inversion-invariant and $I(\cF^{(i)})$ is a distinct component from $\cF^{(i)}$. Because ${I\circ I}$ is the identity map, each contractible component $\cF^{(i)}$ has a unique partner $I(\cF^{(i)})$, and there are an even number of contractible components.

\section{Ons\texorpdfstring{$_d$}{d} in momentum space}\label{momentum space ons d appendix}

In this appendix, we discuss how the ${\mathrm{Ons}_{d}}$ symmetry operators from the main text act on the momentum space fermion operators $c_\k$. In doing so, we will find that ${\mathrm{Ons}_{d}}$ has an interesting subgroup, which is naturally understood in terms of momentum space.

From the symmetry transformations~\eqref{U(1)SymTransf} and~\eqref{maj transl c ops}, the symmetry operators $\ee^{\ii\th Q}$ and $T_{\mb{a}_i}^{(b)}$ act on the momentum space operators as
\begin{align}
    \ee^{\ii\th Q} c_\k \ee^{-\ii\th Q} &= \ee^{-\ii\th} c_\k,\\
    T_{\mb{a}_i}^{(b)} c_\k\, (T_{\mb{a}_i}^{(b)})^{-1} &\!=\! \left(\frac{1 +  \ee^{\ii\k\cdot \mb{a}_i}}2\right)c_\k+ \left(\frac{1 - \ee^{\ii\k\cdot\mb{a}_i}}2\right) c^\dag_{-\k}.
\end{align}
Since the Onsager charges are given by ${Q_{\mb{v}} = T_{\mb{v}}^{(b)} Q\, (T_{\mb{v}}^{(b)})^\dag}$, these expressions can be used to straightforwardly deduce the transformation of $\ee^{\ii\th Q_{\mb{v}}}$ on $c_\k$.

There is a more enlightening way to proceed, however. Let us introduce the vector ${\Psi = \bigoplus_{\k\in\bz^+} \Psi_{\k}}$ where ${\Psi_{\k} = (c_\k,\,c^\dag_{-\k})^\mathsf{T}}$ and $\bz^+$ is a subset of the first Brillouin zone containing only one momentum vector from each pair ${(\k,-\k)}$. The vectors $\Psi$ and $\Psi^\dag$ include every momentum space operator $c_\k$ and $c_\k^\dag$. Furthermore, the symmetry operators $\ee^{\ii\th Q}$ and $T_{\mb{a}_i}^{(b)}$ act on each component $\Psi_{\k}$ of $\Psi$ as
\begin{align}
    \ee^{\ii\th Q} \Psi_\k \ee^{-\ii\th Q} &= \ee^{-\ii\th\si^z}\Psi_\k,\\
    T_{\mb{a}_i}^{(b)}  \Psi_\k\, (T_{\mb{a}_i}^{(b)})^{-1} & = \ee^{\ii\frac{\k\cdot\mb{a}_i}{2}(1-\si^x)}\Psi_\k,
\end{align}
where $\si^x$ and $\si^z$ are Pauli matrices. Using these, we find that $\ee^{\ii\th Q_\mb{v}}$ acts on $\Psi_\k$ as
\begin{equation}
\begin{aligned}
    \ee^{\ii\th Q_\mb{v}} \Psi_\k \ee^{-\ii\th Q_\mb{v}} &= \ee^{-\ii\th(\sin(\k\cdot\mb{v})\,\si^y + \cos(\k\cdot\mb{v})\,\si^z)}\Psi_\k.
\end{aligned}
\end{equation}

The benefit of packaging the momentum space operators into the vector $\Psi$ is that $\ee^{\ii\th Q_\mb{v}}$ acts on $\Psi$ as the matrix $\ee^{\ii\th \mathsf{Q}_\mb{v}}$, where
\begin{equation}
    \mathsf{Q}_\mb{v} = \!\bigoplus_{\k\in\bz^+}\!\mathsf{Q}_{\mb{v}}^{(\k)},\quad \mathsf{Q}_{\mb{v}}^{(\k)} \!= -\sin(\k\!\cdot\!\mb{v})\,\si^y - \cos(\k\!\cdot\!\mb{v})\,\si^z.
\end{equation}
Therefore, the matrices $\mathsf{Q}_\mb{v}$, along with their commutators, provide a faithful representation of the Lie algebra of Ons$_d$.

Using this representation, we find that $\mathrm{Ons}_{d}$ contains the subgroup
\begin{equation}\label{Gk subgroups}
    \prod_{\k\in\bz^+} \!O_\k,
    \qquad
    O_\k = \begin{cases}
        \mathrm{U}(1)\quad &\text{if }2\k\in\La^\vee,\\
        \mathrm{SU}(2)\quad &\text{otherwise.}
    \end{cases}
\end{equation}
The matrices representing each $O_\k$ subgroup are constructed from $\mathsf{Q}_{\mb v}$ using\footnote{The Kronecker delta function $\del_{\mb{q},\k}$ is the delta function on the reciprocal lattice. That means that ${\del_{\mb{q},\k} = 1}$ if ${\mb{q} - \k\in\La^\vee}$ and $0$ otherwise.}
\begin{align}
    \frac{2}{|\La|}\sum_{\mb{v}\in\La} \sin(\mb{k}\cdot\mb{v}) \mathsf{Q}_{\mb{v}} &= \bigoplus_{\q\in\bz^+}
    -(\del_{\mb{q},\k}-\del_{\mb{q},-\k})\,\si^y,\label{sine sums}\\
    \frac{2}{|\La|}\sum_{\mb{v}\in\La} \cos(\mb{k}\cdot\mb{v}) \mathsf{Q}_{\mb{v}} &= \bigoplus_{\q\in\bz^+}
    \,-(\del_{\mb{q},\k}+\del_{\mb{q},-\k})\,\si^z.\label{cosine sums}
\end{align}
When ${\k = -\k}$ modulo reciprocal lattice vectors---equivalently, ${2\k\in\La^\vee}$---the right-hand side of~\eqref{sine sums} is the zero matrix. Therefore, when ${2\k\in\La^\vee}$, the Lie algebra of $O_\k$ is generated by only~\eqref{cosine sums}, which makes the corresponding Lie group U(1). When ${2\k\not\in\La^\vee}$, both~\eqref{sine sums} and ~\eqref{cosine sums} are non-vanishing. The Lie algebra they generate is formed by the Pauli matrices $\si^y$ and $\si^z$, and the corresponding Lie group is SU(2).

\section{The particle number centralizer in \texorpdfstring{$\mathfrak{ons}_d$}{onsd}}\label{max abl alg app}

In this Appendix, we prove that the centralizer of $Q$ in $\mathfrak{ons}_d$, the Lie algebra of Ons$_d$, is spanned by the operators ${\t{Q}_{\mb v} = \frac{1}{2} (Q_{\mb v} + Q_{-\mb v})}$.

We denote a general element of $\mathfrak{ons}_d$ by $Q_{\bm{\al},\bm{\bt}}$. It is decomposed into the basis elements $Q_{\mb v}$ and $G_{\mb v}$ as 
\begin{equation}
    Q_{\bm{\al},\bm{\bt}} = \sum_{{\mb v}\in\La} \al_{\mb v} \, Q_{\mb v} + \sum_{\mb v\in\La^+}\bt_{\mb v} G_{\mb v},
\end{equation}
where $\alpha_{\mb v},\bt_{\mb v}\in\C$ and $\La^+$ is a subset of $d$-dimensional space containing only one lattice vector from each pair ${(\r,-\r)}$. Not every element $Q_{\bm{\al},\bm{\bt}}$ commutes with $Q$. Using the commutation relations~\eqref{Ons d alg}, $Q_{\bm{\al},\bm{\bt}}$ satisfies
\begin{equation}
    [Q, Q_{\bm{\al},\bm{\bt}}] = \ii\!\sum_{\mb{v}\in\La^+}\!\big( 2\, \beta_{\mb v}(Q_{-\mb v} - Q_{\mb{v}}) + (\alpha_{\mb v} - \alpha_{-\mb{v}})G_{\mb v} \big).
\end{equation}
Therefore, $Q_{\bm{\al},\bm{\bt}}$ commutes with $Q$ if and only if ${\alpha_{\mb v} = \alpha_{-\mb v}}$ and ${\bt_{\mb v} = 0}$, and the centralizer of $Q$ is formed by
\begin{equation}
    \sum_{\mb{v}\in\La} \al_{\mb{v} }Q_{\mb v} = \sum_{\mb{v}\in\La} \al_{\mb{v} }\t{Q}_{\mb v}. 
\end{equation}
It is straightforward to show that the charges $\t{Q}_{\mb v}$ mutually commute: ${[\t{Q}_{\mb v_1},\t{Q}_{\mb v_2}]=0}$. Hence, the subalgebra formed by $\t{Q}_{\mb v}$ is abelian, and the centralizer of $Q$ is the maximal abelian subalgebra of $\mathfrak{ons}_d$ containing $Q$.

The commuting charges $\t{Q}_{\mb v}$ satisfy
\begin{align}
    [\t{Q}_{\mb v}, c_\k] &= -\cos (\k \cdot \mb{v}) \: c_\k,\\
    [\t{Q}_{\mb v}, c_\k^\dagger] &= \cos (\k \cdot \mb{v}) \: c_\k^\dagger.
\end{align}
Therefore, the corresponding symmetry operators $\ee^{\ii\lambda \t{Q}_{\mb v}}$ do not mix $c_\k$ with $c_{-\k}^\dagger$, and satisfy
\begin{align}
    \ee^{\ii\la\t{Q}_{\mb v}} c_\k \ee^{-\ii \la \t{Q}_{\mb v}} &=  \: \ee^{-\ii\la \cos (\k \cdot \mb{v}) } c_\k, \\
    \ee^{\ii\la\t{Q}_{\mb v}} c_\k^\dagger \ee^{-\ii \la \t{Q}_{\mb v}} &=  \: \ee^{\ii\la \cos (\k \cdot \mb{v}) } c_\k^\dagger.
\end{align}

\section{Magnetic Ons\texorpdfstring{$_2$}{2} symmetry in the \texorpdfstring{$\pi$}{pi}-flux model}
\label{magnetic app}

In this appendix, we consider a magnetic $\mathfrak{ons}_2$ algebra generated by acting magnetic Majorana translations on $Q$ in the presence of a uniform magnetic field. A similar analysis of the Onsager algebra in ${(3+1)}$D staggered fermion systems was performed in~\cite{OY250904906, AKT251106198}.

Consider a nearest-neighbor tight-binding Hamiltonian with a uniform magnetic flux $\varphi$. For simplicity, we will restrict our discussion to the ${d=2}$ square lattice with primitive lattice vectors ${\mb{a}_1 = \hat{x} \equiv (1,0)}$ and ${\mb{a}_2 = \hat{y} \equiv (0,1)}$. We assume that ${L_x , L_y = 0 \bmod 4}$. Defining the index ${\mu=1, 2}$ and denoting the magnetic flux by $\varphi$, we consider the Hamiltonian
\begin{equation}\label{fluxTBmodel}
		H_{\varphi} = \sum_{\r,\mu} \left(\ii c_{\r}^\dag c_{\r+\mb a_\mu} \ee^{\ii A_{\r,\mu}} + \mathrm{hc}\right),
        \quad 
        A_{\r,\mu} = -\frac{\varphi
		}{2}\eps_{\mu\nu}r_\nu.
\end{equation}
We will refer to this as the $\varphi$-flux model. When ${\varphi=\pi}$, this model is known as the staggered fermion model~\cite{PhysRevD.16.3031} in the high-energy literature. Its continuum limit is described by two copies of the massless Dirac fermion field theory.

When ${\varphi\neq 0 \bmod 2\pi}$, the Hamiltonian~\eqref{fluxTBmodel} does not commute with $T_{\mb v}^{(b)}$. A natural question is whether an alternative set of Majorana operators exists whose Majorana translations do commute with $H_\varphi$. As we now explain, this is only possible when ${\varphi = 0,\pi\bmod 2\pi}$. Indeed, consider the most general decomposition of the complex fermion operator $c_\r$ into real fermion operators $a_\r$ and $b_\r$: ${c_\r = \frac{1}{2} (f(\r)\, a_\r + \ii g(\r)\, b_\r)}$. In order for $c_\r$ to satisfy the canonical anticommutation relations, the coefficients ${f(\r) = \pm g(\r) = \ee^{\ii\th(\r)}}$. Plugging ${c_\r = \ee^{\ii \theta(\r)}\left( \frac{ a_\r \pm \ii b_\r}{2} \right)}$ into the $\varphi$-flux Hamiltonian~\eqref{fluxTBmodel}, Majorana translations are conserved only when ${\varphi= 0,\pi \bmod 2\pi}$, as claimed. Because the case of $\varphi = 0 \bmod 2\pi$ was covered in the main text, we will focus on ${\varphi= \pi \bmod 2\pi}$ here.

For ${\varphi = \pi}$, we consider Majorana decomposition 
\begin{equation}
    c_\r = \ee^{\ii \frac{\pi}{2} r_x r_y} \left(\frac{a_\r + \ii b_\r}{2}\right).
\end{equation}
Because we assume ${L_x , L_y = 0 \bmod 4}$, the phase $\ee^{\ii \frac{\pi}{2} r_x r_y} $ is single-valued under the periodic boundary conditions. The $\pi$-flux Hamiltonian in terms of these Majorana operators is 
\begin{equation}\label{pi flux Ham Maj}
	H_{\pi} = \frac{\ii}{2}\sum_{\r} [ a_\r a_{\r+\hat x} + b_\r b_{\r+\hat x} + (-1)^{r_x} (a_\r a_{\r+\hat y} + b_\r b_{\r+\hat y}) ],
\end{equation}
which is two copies of the $\pi$-flux Majorana fermion model from Refs.~\cite{PhysRevB.73.201303,Liu:2015dgq,Affleck:2017ubr,SZ260101191}. This Hamiltonian commutes with the (modified) Majorana translation operators\footnote{We follow the convention that, for example, $\prod_{i=1}^{N} b_{\bm{v}_i}  = b_{\bm{v}_1}b_{\bm{v}_2}\cdots b_{\bm{v}_N}$.}
\begin{equation}
	T^{(b,\pi)}_{\r} = \bigg(\prod_{v_y=1}^{L_y}\prod_{v_x=1}^{L_x} b_{\bm{v}}^{v_y\,r_x} \bigg)\, T^{(b)}_{\r}.
\end{equation}
We refer to these operators as the magnetic Majorana translation operators because they satisfy the following algebra:
\begin{equation}
	T^{(b,\pi)}_{\r} T^{(b,\pi)}_{\t{\r}} = P_b^{r_y \t{r}_x - r_x \t{r}_y}\,
	T^{(b,\pi)}_{\t{\r}}
	T^{(b,\pi)}_{\r}.
\end{equation}
Here, ${P_b = \prod_{v_y=1}^{L_y}\prod_{v_x=1}^{L_x} b_{\bm{v}}}$ is the fermion parity for the $b$-Majorana fermion, which acts on the fermions as
\begin{equation}
P_b a_\r P_b^{-1} ,~~~
P_b b_\r P_b^{-1} = -b_\r.
\end{equation}
The magnetic Majorana translations act on the fermions as
\begin{equation}
\begin{aligned}
 T^{(b,\pi)}_{\mb v} a_\r (T^{(b,\pi)}_{\mb v})^\dag &= a_\r,\\
 T^{(b,\pi)}_{\mb v} b_\r (T^{(b,\pi)}_{\mb v})^\dag &= (-1)^{r_y v_x}b_{\r+\mb v}.
\end{aligned}
\end{equation}
Therefore, the $\pi$-flux Majorana translation symmetry group forms an extension of the $0$-flux Majorana translation group by the $\Z_2$ Majorana number parity symmetry group.

Since ${[H_\pi, Q] = [H_\pi, T^{(b,\pi)}_{\r}]=0}$, we can construct a set of U$(1)$ charges by acting the ${T^{(b,\pi)}_{\r}}$ translation operators on the fermion number operator $Q$, which yields
\begin{equation}
\begin{aligned}
	Q_{\r}^\pi &= (-1)^{r_x r_y}\, T^{(b,\pi)}_{\r}  Q \left(T^{(b,\pi)}_{\r} \right)^\dag,\\
    &= 
	\frac{\ii}2 \sum_{\mb{v}} (-1)^{r_x v_y} a_{\mb{v}} b_{\mb{v}+\r}.
\end{aligned}
\end{equation}
The multiplicative factor ${(-1)^{r_x r_y}}$ is added for convenience. These charge operators $Q_{\r}^\pi$ are conserved and satisfy ${[H_\pi, Q^\pi_\r] = 0}$.

The charge operators $Q_{\r}^\pi$ do not satisfy the $\mathfrak{ons}_2$ algebra, but instead satisfy a modified version. Let us define 
\begin{align}
A_{\r}^\pi &= \frac{\ii}{2}\sum_{\mb v} \big(
(-1)^{r_x v_y}a_{\mb{v}}a_{\mb{v}+\r} - \delta_{\r,0} 
\big),\\
B_{\r}^\pi &= -\frac{\ii}{2}\sum_{\mb v} \big((-1)^{r_x v_y} b_{\mb{v}}b_{\mb{v}+\r} -\delta_{\r,0}\big).
\end{align}
The shifts by $\delta_{\r,0}$ are included to set ${A_0^\pi = B_0^\pi = 0}$ and ensure that $A_\r^\pi$ and $B_\r^\pi$ are Hermitian. The charges $Q^\pi_\r$ satisfy
\begin{equation*}
    [Q_{\r}^\pi, Q_{\r'}^\pi] = \!\ii  \left((-1)^{r_x (r_y + r_y')} A_{\r'-\r}^\pi \!+\! (-1)^{(r_x + r_x' ) r_y} B_{\r' - \r}^\pi\right),
\end{equation*}
and the $A_{\r}^\pi$ and $B_{\r}^\pi$ operators satisfy
\begin{align*}
	&[A_{\r}^\pi, \!A_{\r'}^\pi]  \!=\! \ii\! \left(\!(-1)^{r_x' r_y} \!-\! (-1)^{r_x r_y'}\!\right)\! \!\left(A_{\r + \r'}^\pi \!-\! (-1)^{r_x r_y} A_{\r' - \r}^\pi\right)\!,
    \\
	&[B_{\r}^\pi, \!B_{\r'}^\pi]  \!= \!\ii\! \left(\!(-1)^{r_x r_y'} \!-\! (-1)^{r_x' r_y}\!\right)\!\! \left(B_{\r + \r'}^\pi\! -\! (-1)^{r_x r_y} B_{\r' - \r}^\pi\right)\!,
    \\
	&[A_{\r}^\pi, B_{\r'}^\pi] = 0,
    \\
	&[Q_{\r}^\pi, A_{\r'}^\pi] =  \ii (-1)^{r_x r_y'} \left((-1)^{r_x' r_y'} Q_{\r- \r'}^\pi - Q_{\r + \r'}^\pi\right) ,
    \\
	&[Q_{\r}^\pi, B_{\r'}^\pi] =  \ii (-1)^{r_x' r_y} \left((-1)^{r_x' r_y'} Q_{\r- \r'}^\pi - Q_{\r + \r'}^\pi\right).
\end{align*}
We refer to the algebra generated by $Q^\pi_\r$ as the magnetic $\mathfrak{ons}_2$, which follows from these commutation relations.

We now derive the most general local symmetric Hamiltonian that commutes with magnetic Majorana translations and $Q$. In order to commute with ${T^{(b,\pi)}_{\mb v}}$ and remain local, the $a$ and $b$ Majorana operators must decouple in the Hamiltonian. Furthermore, following the argument from~\cite{CPS240912220,PCS241218606} and the main text, to commute with $Q$, the Hamiltonian must be quadratic. The most general local, quadratic Hamiltonian commuting with $Q$ and ${T^{(b,\pi)}_{\mb v}}$ is
\begin{equation}\label{general pi flux model}
    H =  \ii \sum_{\r, \mb v} (-1)^{v_y r_x}  f_{\mb v} \left(a_\r a_{\r+\mb v} + b_\r b_{\r+\mb v} \right).
\end{equation}
The coefficients $f_{\mb v}$ are real. We recover the $\pi$-flux Hamiltonian~\eqref{pi flux Ham Maj} upon setting ${f_{\hat x} = f_{\hat y} = \frac{1}{2}}$ and zero otherwise. 

In contrast to its $\mathrm{Ons}_2$ counterpart, the magnetic $\mathrm{Ons}_2$ symmetry does not enforce a Fermi surface. Indeed, the $\pi$-flux Hamiltonian~\eqref{pi flux Ham Maj} is gapless with two Dirac cones, but does not have a Fermi surface.
In fact, this magnetic $\mathrm{Ons}_2$ symmetry is compatible with a trivially gapped phase and, therefore, anomaly-free.\footnote{We thank Ioanna Kourkoulou and David Tong for pointing out an error regarding this point made in a previous version of this manuscript.} For example, the Hamiltonian
\begin{equation}
    \ii \sum_{\r} (-1)^{r_x} (a_\r a_{\bm{\r + \hat{x}+\hat{y}}}
    +
    b_\r b_{\bm{\r + \hat{x}+\hat{y}}})
\end{equation}
commutes with $Q$ and ${T^{(b,\pi)}_{\mb v}}$ and has a unique gapped ground state.


\def\bibsection{\section*{\refname}}

 \bibliographystyle{ytphys}
\bibliography{refs}

\end{document}

%% file: sdpPreamble.tex
\usepackage{etoolbox}
\makeatletter
\patchcmd{\@author@present@script}
  {\gdef\comma@space{\textsuperscript{,\,}}}
  {\gdef\comma@space{\textsuperscript{\,}}}
  {}{}
\makeatother

\usepackage{graphicx}
\usepackage{dcolumn}
\usepackage{bm}
\usepackage{multirow}

\usepackage{centernot}

\usepackage[normalem]{ulem}
\usepackage{cancel}

\usepackage{amsmath}
\usepackage{amsthm}
\usepackage{amstext}
\usepackage{amssymb}
\usepackage{mathrsfs}
\usepackage{amsfonts}
\usepackage{amsbsy} 
\usepackage{tensor}
\usepackage{physics}

\usepackage{wasysym}

\usepackage{latexsym}
\usepackage[american]{babel}
\usepackage{bbm}
\usepackage[colorlinks=true, citecolor=blue!90!black, linkcolor=blue!90!black, linktocpage=true, urlcolor=red!70!black]{hyperref}

\newcommand*{\figref}[2][]{%
  \hyperref[{#2}]{%
    \ref*{#2}%
    \ifx\\#1\\%
    \else
      \,#1%
    \fi
  }%
}

\usepackage[all,matrix,cmtip]{xy}

\usepackage{csquotes}
\MakeOuterQuote{"}

\usepackage{color}
\definecolor{red}{rgb}{1,0,0}
\definecolor{blue}{rgb}{0,0,1}
\definecolor{dblue}{rgb}{0,0,0.4}
\definecolor{green}{rgb}{0,1,0}
\definecolor{black}{rgb}{0,0,0}
\definecolor{white}{rgb}{1,1,1}
\definecolor{niceBlue}{RGB}{20,10,237}

\usepackage[dvipsnames]{xcolor}

\definecolor{brn}{rgb}{.8,.4,.0}
\definecolor{redo}{rgb}{1,.5,.0}
\definecolor{ddgrn}{rgb}{0,0.4,0}
\definecolor{dgrn}{rgb}{0,0.55,0}
\definecolor{dbl}{rgb}{0,0,0.5}

\usepackage[bbgreekl]{mathbbol}

\newcommand{\Z}{\mathbb{Z}}
\newcommand{\C}{\mathbb{C}}
\newcommand{\R}{\mathbb{R}}

\newcommand{\p}[1]{\prime\,}

\renewcommand{\t}[1]{\widetilde{#1}}

\newcommand{\ii}{\hspace{1pt}\mathrm{i}\hspace{1pt}}
\newcommand{\ee}{\hspace{1pt}\mathrm{e}}

\newcommand{\<}{\langle}
\renewcommand{\>}{\rangle}

\newcommand{\pp}{\partial}

\newcommand{\bpm}{\begin{pmatrix}}
\newcommand{\epm}{\end{pmatrix}}
\newcommand{\bmm}{\begin{matrix}}
\newcommand{\emm}{\end{matrix}}

\newcommand{\cF}{ {\cal F} }

\newcommand{\cO}{ {\cal O} }

\usepackage{euscript}

\newcommand{\al}{\alpha} 
\newcommand{\bt}{\beta} 
\newcommand{\del}{\delta} 
 
\newcommand{\eps}{\epsilon}

\newcommand{\la}{\lambda} 
\newcommand{\La}{\Lambda}

\renewcommand{\th}{\theta} 
 
\newcommand{\si}{\sigma}

\usepackage{tikz}
\usepackage{tikz-cd}
\usetikzlibrary{arrows}
\usetikzlibrary{intersections}
\usetikzlibrary{shapes.geometric}
\usetikzlibrary{decorations.pathmorphing, patterns,shapes}
\usetikzlibrary{decorations.markings}

\usepackage{enumerate}

\usepackage{amsthm}

\theoremstyle{definition}

\usetikzlibrary{calc}

\usepackage{slashed}

\usepackage{booktabs}

\newcommand{\ie}{\begin{equation}\begin{aligned}[]}
\newcommand{\fe}{\end{aligned}\end{equation}}

\usepackage{comment}
\usepackage{mathtools}

\renewcommand{\t}[1]{\widetilde{#1}}
\renewcommand{\k}{\mathbf{k}}

\newcommand{\q}{\mathbf{q}}
\renewcommand{\p}{\mathbf{p}}
\renewcommand{\r}{\mathbf{r}}

\newcommand{\bz}{{\rm BZ}}

\newcommand{\mb}[1]{\mathbf{#1}}
\usepackage{braket}